\documentclass{elsart}

\usepackage{natbib}

\usepackage{graphicx}
\begin{document}

\begin{frontmatter}

% Title, authors and addresses

% use the thanksref command within \title, \author or \address for footnotes;
% use the corauthref command within \author for corresponding author footnotes;
% use the ead command for the email address,
% and the form \ead[url] for the home page:
% \title{Title\thanksref{label1}}
% \thanks[label1]{}
% \author{Name\corauthref{cor1}\thanksref{label2}}
% \ead{email address}
% \ead[url]{home page}
% \thanks[label2]{}
% \corauth[cor1]{}
% \address{Address\thanksref{label3}}
% \thanks[label3]{}

\title{Turbulence Generation by Substructure Motion in Clusters of Galaxies}

% use optional labels to link authors explicitly to addresses:
% \author[label1,label2]{}
% \address[label1]{}
% \address[label2]{}

\author{M. Takizawa}

\address{Department of Physics, Yamagata University, Kojirakawa-machi
 1-4-12, Yamagata 990-8560, Japan}

\begin{abstract}
Clusters of galaxies form through major merger and/or absorption
of smaller groups. In fact, some characteristic structures such as cold fronts,
which are likely relevant to moving substructures, are found by 
{\it Chandra}. 
It is expected that moving substructures generate turbulence in the
intracluster medium (ICM). Such turbulence probably plays a crucial role
in mixture and transport of gas energy and heavy elements, and
particle acceleration. The {\it Astro-E2} satellite, which is planned to be
launched in 2005, will detect broadened lines due to turbulent motion.
In order to explore the above-mentioned issues, it is important to investigate
the generation processes and structure of ICM turbulence.
We investigate the ICM dynamical evolution in and around a moving substructure
with three-dimensional hydrodynamical simulations. Eddy-like structures
develop near the boundary between the substructure and the ambient ICM through
Kelvin-Helmholtz instabilities. 
Because of these structures, characteristic patterns appear 
in the line-of-sight velocity distribution of the ICM.
\end{abstract}


\ead{takizawa@sci.kj.yamagata-u.ac.jp}

\end{frontmatter}

% main text
\section{Introduction}

Clusters of galaxies are the largest virialized objects in the present
Universe. According to the standard scenario of structure formation
in the Universe, larger objects form more recently. Thus, clusters of
galaxies are the virialized objects which form in the most recent epoch
in the Universe. In fact, some clusters are forming now, which is
evident from the moving substructures that are found through X-ray
observations (e.g. Markevitch et al., 2000, 2002; Vikhlinin et al., 2001).

A moving substructure causes various characteristic structures in the
intracluster medium (ICM). A bow shock and a contact discontinuity will
form in front of it. The latter most likely corresponds to a ``cold
front'' which is found by {\it Chandra} in a number of clusters 
(e.g. Markevitch et al., 2000; Vikhlinin et al., 2001). Moving
substructures generate turbulence in the ICM through fluid instabilities
\citep{Hein03}.
Turbulence probably plays an important role in cluster evolution. It may
have a significant impact on the transport and mixture of heavy elements 
and on those of thermal energy. 
Fluid turbulence probably generates magnetic turbulence,
which accelerates non-thermal particles and causes various high energy
phenomena in the intracluster space (e.g. Ohno et al., 2002; Fujita et
al., 2003; Brunetti et al., 2004). The {\it Astro-E2} satellite 
will detect broadened lines due to
turbulent motion \citep{Suny03}. 
In order to study the above-mentioned issues, it is
crucial to clarify the generation processes and structure of the ICM
turbulence.

\section{The Simulations}

In the present study, we used the Roe TVD scheme to follow the dynamical
evolution of the ICM (see Hirsch, 1990). The Roe scheme is a Godunov-type
method and is based on a linearized Riemann solver \citep{Roe81}. 
It is relatively simple and good at capturing shocks without any
artificial viscosity. Using the MUSCLE approach 
and a minmod TVD limiter, we obtained second-order accuracy without any
numerical oscillations around discontinuities. To avoid negative
pressure, we solved the equations for the total energy and entropy
conservation simultaneously. This method is often used in astrophysical
hydrodynamical simulations where high Mach number flows
can occur \citep{Ryu93}.

Our initial conditions are as follows.  We set a subcluster 
in the center of the simulation box. The subcluster's gravitational
potential is represented by that of a King distribution, whose total mass
and core radius are $10^{14} M_{\odot}$ and $100$ kpc, respectively. 
The subcluster's radius is $500$ kpc. Inside the subcluster, the gas is
assumed to be in hydrostatic equilibrium with an isothermal temperature
distribution. We set the initial temperature inside the subcluster so
that $\beta_{\rm spec}=0.8$. We assume that the mass density of the gas
is a tenth of that of the DM in the center. Outside the subcluster, the gas
pressure is the same as that on the outer boundary of the subcluster and
the density is half of the density on that boundary.
Only in front of the subcluster we set the initial velocity to 
twice the sound speed. 
The size of the simulation box and the number of grid points
are (2 Mpc)$^3$ and $200^3$, respectively. We set constant
inflow boundary conditions only for the front boundary. Free boundary
conditions are adopted for the other boundaries.

\section{Results}

\begin{figure}
 \begin{center}
   \includegraphics*[scale=0.85]{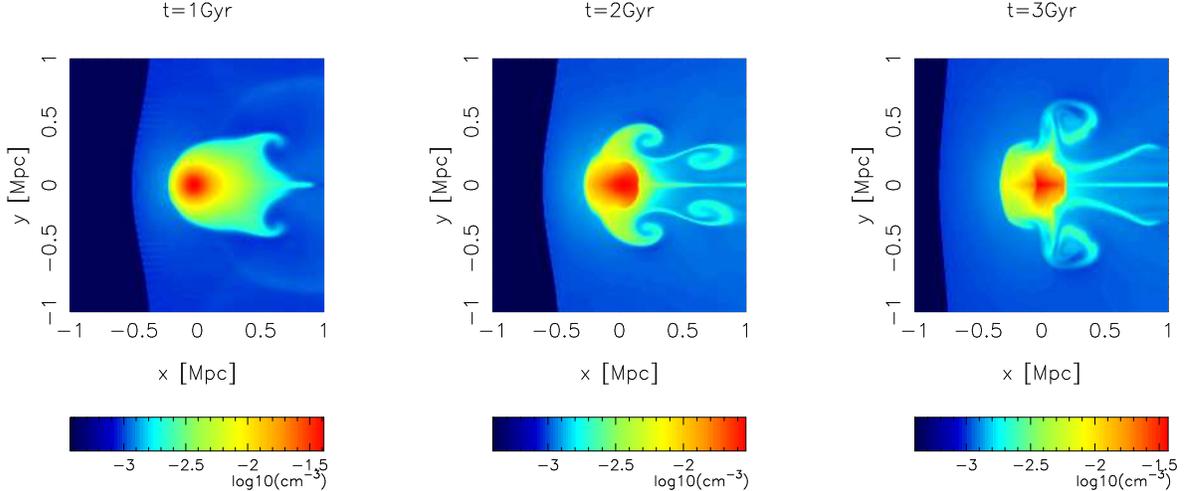}
 \end{center}
 \caption{Snapshots of the density distribution on the $z=0$ surface at $t=1,2,$
          and $3$ Gyr. The substructure's gas in the outer region is
          stripped off behind the shock because of the ram pressure. 
          A Kelvin-Helmholtz instability developed on the boundary between the
          substructure and the ambient gas, and prominent eddy-like 
          structures form just behind the substructure.}
 \label{fig:density}
\end{figure}

Figure \ref{fig:density} shows snapshots of the density distribution on
the $z=0$ surface at $t=1,2,$ and 3 Gyr. Two shocks propagate forward and
backward. The shock which propagates backward has left the simulation box
before $t=1$ Gyr. Thus, only the forward shock is seen in figure
\ref{fig:density}. The substructure's gas in the outer region is
stripped off behind the shock because of the ram pressure. In addition,
a Kelvin-Helmholtz instability developed on the boundary between the
substructure and the ambient gas, and prominent eddy-like structures
form just behind the substructure.

\begin{figure}
 \begin{center}
   \includegraphics*[scale=0.85]{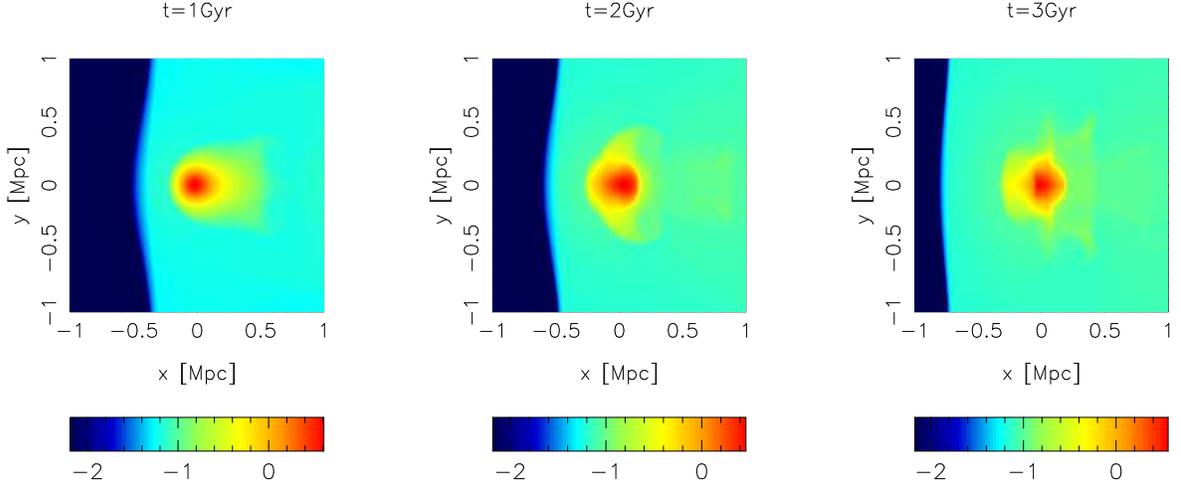}
 \end{center}
 \caption{Snapshots of the X-ray surface brightness distribution at $t=1,2,$
          and $3$ Gyr. The line-of-sight is just along the $z$-axis. 
          Eddy-like structures around the subcluster become less clear 
          because of contamination of the foreground and background 
          components. }
 \label{fig:xrsb}
\end{figure}

Although the eddy-like structures are clearly seen in figure \ref{fig:density},
In actual observations we can get only information integrated along the
line-of-sight. Therefore, these structures may be less clear in
observed quantities. Figure \ref{fig:xrsb} shows snapshots of the X-ray
surface distribution seen along the $z$-axis. We assume that the volume
emissivity is proportional to $\rho^2 T^{1/2}$, where $\rho$ and $T$ are
the gas density and temperature, respectively. Because of contamination of
the foreground and background components, the eddy-like structures around
the subcluster become less clear. 

\begin{figure}
 \begin{center}
   \includegraphics*[scale=0.85]{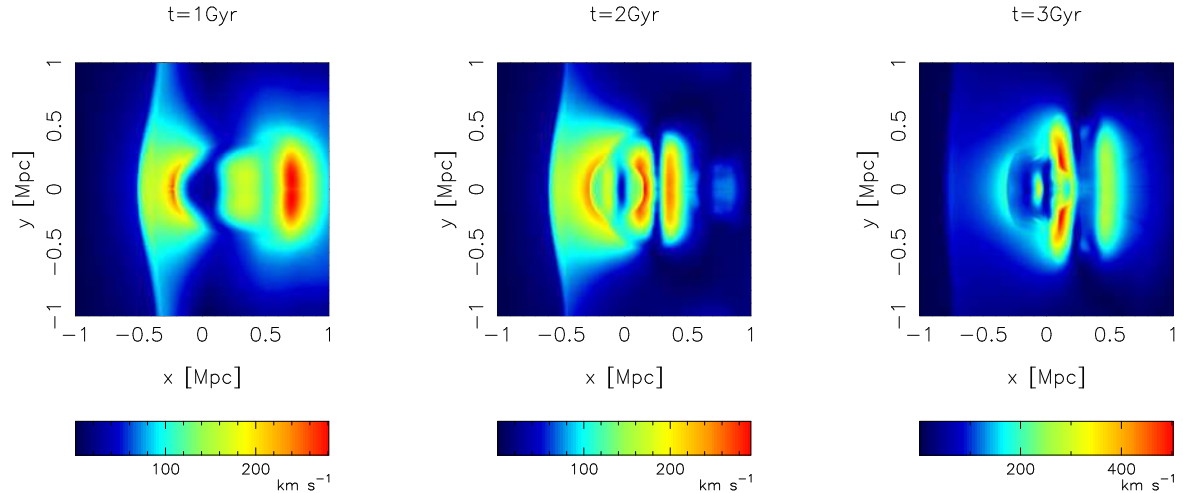}
 \end{center}
 \caption{Standard deviation distribution of the line-of-sight velocity
          component at $t=1,2,$ and $3$ Gyr. Around the contact
          discontinuity in front of the substructure, the root mean square
          velocity is $\sim 200$ km s$^{-1}$ because of the bulk flows
          along the convex boundary. Behind the substructure, 
          on the other hand, it becomes $300 \sim 500$ km s$^{-1}$
          because of the turbulent motion associated with the eddies. }
 \label{fig:rms_vz}
\end{figure}

The {\it Astro-E2} satellite has a high
resolution X-ray spectrometer XRS. It will be able to detect 
$\sim 100$ km s$^{-1}$ gas flows through Doppler broadening and/or shift
of the emission lines. Because of the symmetry, no line shift is expected
in our simulations when the line-of-sight is along the $z$-axis. 
On the other hand,
line broadening is expected because of the bulk flows and turbulent
motion. The Line-of-sight velocity dispersion (or standard deviation,
equivalently) is a good estimator of line-broadening. 
Figure \ref{fig:rms_vz} shows the emissivity-weighted 
standard deviation distribution of the line-of-sight velocity component. 
Around the contact discontinuity in front of the substructure, 
the root mean square velocity is $\sim 200$ km s$^{-1}$ because of the 
bulk flows along the convex boundary. Behind the substructure, 
on the other hand, 
it becomes $300 \sim 500$ km s$^{-1}$ because of the turbulent motion
associated with the eddies. 

\section{Summary}

We investigate the gas motion in and around a moving substructure in a
cluster of galaxies using a three dimensional Roe-TVD hydrodynamic code. 
After the ICM in the outer region of the substructure is stripped off
because of the ram pressure, a Kelvin-Helmholtz instability develops on
the boundary between the substructure and the ambient ICM.
Then, eddy-like structures form there and the cold gas clouds 
originating from
the substructure fly backwards. The line-of-sight velocity dispersion of
the ICM becomes large around and in the back of the substructure
remnant when we see the system from the direction perpendicular to the
substructure motion. The root mean square velocity becomes 
$300 \sim 500$ km s$^{-1}$, which is large enough to be detected by 
the {\it Astro-E2} XRS.

Although in this paper we only showed the results in a rather simple
case where a subcluster is moving at a constant velocity in a
uniform background,
we are planning to investigate more realistic situations
where the ambient ICM properties can vary with time.
For instance, simulations of a subcluster infalling into a larger
cluster and sloshing near the cluster center will be useful.

In this work, we neglected the magnetic field in the ICM. However, it 
is possible that the magnetic field plays a crucial role to suppress the
development of Kelvin-Helmholtz instabilities \citep{Asai03}. On the
other hand, the evolution and structure of MHD turbulence contains important
information about particle acceleration in the ICM.
Therefore, high resolution MHD simulations will be important as future work.

The numerical computations were carried out on the VPP5000 at
the Astronomical Data Analysis Center of the National Astronomical
Observatory, Japan, which is an inter-university research institute of
astronomy operated by the Ministry of Education, Science, Culture, and
Sports. M. T. was supported in part by a Grant-in-Aid from the Ministry
of Education, Science, Sports, and Culture of Japan (16740105).

% The Appendices part is started with the command \appendix;
% appendix sections are then done as normal sections
% \appendix

% \section{}
% \label{}

% Bibliographic references with the natbib package:
% Parenthetical: \citep{Bai92} produces (Bailyn 1992).
% Textual: \citet{Bai95} produces Bailyn et al. (1995).
% An affix and part of a reference:
%   \citep[e.g.][Ch. 2]{Bar76}
%   produces (e.g. Barnes et al. 1976, Ch. 2).

\end{document}